\documentclass[prd,aps,showpacs,tightenlines,nofootinbib]{revtex4}
\usepackage{graphicx}

\newcommand{\CL}{{\cal L}}

\newcommand{\bear}{\begin{array}}  \newcommand{\eear}{\end{array}}
\newcommand{\bea}{\begin{eqnarray}}  \newcommand{\eea}{\end{eqnarray}}
\newcommand{\beq}{\begin{equation}}  \newcommand{\eeq}{\end{equation}}
\newcommand{\bef}{\begin{figure}}  \newcommand{\eef}{\end{figure}}
\newcommand{\bec}{\begin{center}}  \newcommand{\eec}{\end{center}}
\newcommand{\bed}{\begin{description}}  \newcommand{\eed}{\end{description}}
\newcommand{\non}{\nonumber}  
\newcommand{\lmk}{\left(}  \newcommand{\rmk}{\right)}
  
\newcommand{\lhk}{\left \{ }  \newcommand{\rhk}{\right \} }
\newcommand{\del}{\partial}  

\newcommand{\bib}{\bibitem} 
\newcommand{\la}{\left\langle} \newcommand{\ra}{\right\rangle}

\def\NPB#1#2#3{Nucl. Phys. {\bf B#1}, #2 (19#3)}

\def\PLB#1#2#3{Phys. Lett. B {\bf #1}, #2 (19#3)}

\def\PRDD#1#2#3{Phys. Rev. D {\bf #1}, #2 (20#3)}

\def\PRL#1#2#3{Phys. Rev. Lett. {\bf#1}, #2 (19#3)}

\newcommand{\lesssim}{ \mathop{}_{\textstyle \sim}^{\textstyle <} }

\begin{document}

\title{Spontaneous baryogenesis in warm inflation}
\author{Robert H. Brandenberger}
\affiliation{Physics Department, Brown University, Providence, RI
 02912, USA} 
\author{Masahide Yamaguchi}
\affiliation{Physics Department, Brown University, Providence, Rhode
Island 02912 USA}
\date{\today}
\begin{abstract}
  We discuss spontaneous baryogenesis in the warm inflation scenario.
  In contrast with standard inflation models, radiation always exists
  in the warm inflation scenario, and the inflaton must be directly
  coupled to it. Also, the transition to the post-inflationary
  radiation dominated phase is smooth and the entropy is not
  significantly increased at the end of the period of inflation. In
  addition, after the period of warm inflation ends, the inflaton does
  not oscillate coherently but slowly rolls. We show that as a
  consequence of these features of warm inflation, the scenario can
  well accommodate the spontaneous baryogenesis mechanism, provided
  that the decoupling temperature $T_{D}$ of the baryon or the $B-L$
  violating interactions is higher than the temperature of radiation
  during the late stages of inflation.
\end{abstract}

\pacs{98.80.Cq \hspace{7.9cm} BROWN-HET-1345} \maketitle


\section{Introduction}

\label{sec:introduction}

Inflation gives the most natural solution to some of the problems of
standard big bang cosmology such as the horizon problem and the
flatness problem, and provides a causal mechanism for the origin of
the primordial density perturbations whose present state is being
mapped to high precision by observational cosmologists
\cite{inflation}.  There are two types of inflation models. The first
(to which most of the proposed inflation models belong) is isentropic:
any preexisting radiation before the onset of inflation is completely
diluted away during inflation, and the radiation must then be
regenerated at the end of the phase of inflation during inflationary
reheating.  The other is nonisentropic and called warm inflation
\cite{warm} (see also \cite{prewarm}): here radiation is continuously
produced by the decay of the inflaton, the scalar field which
generates inflation, and this decay in turn supports the slow-roll
behavior of the inflaton. In this scenario, the temperature of
radiation remains large during the period of inflation, and no
nonadiabatic radiation generation mechanism needs to be postulated at
the end of inflation.

There are significant differences between the warm inflation scenario
and standard isentropic inflation. Most importantly for the purpose
explored in this paper, the inflaton should be coupled to ordinary
matter, whereas in standard inflation it is usually assumed to be a
gauge singlet. Also, since radiation always exists, the transition
from the inflationary period to the radiation dominated period is
straightforward. In particular, after the end of the inflationary
phase the inflaton field will in general still be rolling slowly,
rather than oscillating about the minimum of its potential as happens
in the standard inflationary models. In addition, in warm inflation
the primordial density fluctuations originate from thermal
fluctuations rather than quantum fluctuations of the inflaton
\cite{thermal}.

A requirement for warm inflation is that the constant rate which
describes the decay of the inflaton into particles dominates over the
Hubble damping coefficient in the inflaton equation of motion. It is a
nontrivial problem to obtain such a large decay width and realize warm
inflation. The dynamics of warm inflation has been investigated in the
context of quantum field theory \cite{BGR,YL}, but the dissipative
dynamics is not yet understood fully. It has been suggested that, in
order to realize warm inflation, the inflaton must couple to a very
large number of particle species \cite{YL}. Some models which achieve
this from first principles were proposed \cite{principle}.

We will show that as a consequence of the above-mentioned differences
between warm inflation and standard inflation, it is easier to obtain
spontaneous baryogenesis in the context of warm inflation than in the
context of standard inflation. Most importantly, the inflaton must be
coupled to ordinary matter in a warm inflation scenario, and thus it
is natural to assume that it is not a gauge singlet, whereas in
standard inflation it is usually assumed to be a gauge singlet.  In
addition, in the standard isentropic inflation scenario, the inflaton
oscillates coherently after inflation. Due to the friction term caused
by the current violating operator, this oscillation becomes asymmetric
so that a baryon charge or a $B-L$ charge is generated. This mechanism
also applies to the oscillation of a Nambu-Goldstone boson like an
axion and has been discussed in detail in Ref. \cite{DF}. As shown in
that reference, the oscillations lead to a suppression of the strength
of net baryogenesis over what would be obtained using the naive
classical analysis. However, in warm inflation, the inflaton does not
oscillate coherently but continues to slowly roll even after inflation
ends.  Thus, the analysis of spontaneous lepto/baryogenesis in the
warm inflation scenario will be different. As we will show, the
classical analysis is justified in this case and hence baryogenesis
will be more efficient.

We shall assume that among the many particles the inflaton $\phi$
couples to, it will also couple - albeit derivatively - to the baryon
current or to the $B-L$ current.  Following the arguments by Cohen and
Kaplan \cite{CK}, we will show that baryo/leptogenesis may be possible
if the inflaton has a derivative coupling to such a current given by
\beq
  \CL_{\rm eff} = \frac{1}{M}\,\del_{\mu}\phi\,J^{\mu},
\eeq
where $M$ is the cutoff scale which describes the physics of baryon
number violation. In the above, $J^{\mu}$ is either the baryon current
or the $B-L$ current. Integrating this coupling by parts, we have an
interaction term given by
\beq
  \CL_{\rm eff} = - \frac{1}{M}\,\phi\,\del_{\mu}J^{\mu}.
\eeq
If baryon number conservation or $B-L$ number conservation is
violated, the divergence does not disappear and is replaced by a
current violating operator, which can cause baryo/leptogenesis like in
Affleck-Dine baryogenesis \cite{AD}.

In this paper, we explore in detail the spontaneous baryo/leptogenesis
mechanism in warm inflation. In the next section, we briefly review
the warm inflation scenario. In Sec III, we discuss the possibility of
spontaneous baryo/leptogenesis in warm inflation. In the final
section, we summarize our results.

\section{Dynamics of warm inflation}

\label{sec:warm}

In this section, we briefly review the dynamics of warm inflation.
First of all, we assume that the inflaton couples to a large number of
particles, generating a large and time-independent dissipative
constant $\Gamma$ (related to the decay width). Then, the equation of
motion of the inflaton and the time development of the energy density
of radiation $\rho_{\rm r}$ are given by\footnote{Strictly speaking,
as shown in \cite{BGR,YL,principle}, the dissipative coefficient
$\Gamma$ has a more complicated form and is in fact nonlocal in
time. However, as shown in [4-6], by using the Markovian adiabatic
approximation, the dissipative term becomes local in time and has a
rather simple form in the high temperature limit. Generally speaking,
the dissipative coefficient still depends on the field value. But, if
we consider a Yukawa interaction and the self-energy contribution, for
example, the dissipative coefficient becomes proportional to the
temperature, which is a constant in the context of this paper. Thus,
as a first approximation, we assume that the dissipative term is local
in time, proportional to the time derivative of the field, and in
particular that its coefficient is a constant.}
\bea
 \ddot{\phi} + (3H + \Gamma)\dot{\phi} + V'(\phi) = 0, && 
 \label{eq:eqn} \\
 \dot{\rho_{\rm r}} + 4H\rho_{\rm r} = \Gamma \dot{\phi}^2 &&
 \label{eq:reqn} 
\eea
with the potential given by $V(\phi) = \frac12 m^2 \phi^2$. Here, the
derivative of the potential is taken with respect to $\phi$, and $H$
is the Hubble parameter given by
\beq
  H^2 = \frac{1}{3M_{G}^2}(\rho_{\phi} + \rho_{\rm r}),
\eeq
with $\rho_{\phi}$ being the energy density of the inflaton and $M_{G}
\simeq 2.4 \times 10^{18}$ GeV denoting the reduced Planck scale.

For successful warm inflation, we require the following five
conditions:
\bed
  \item[(i)]  $\rho_{\rm r} \ll V$,
  \item[(ii)] ${1 \over 2} {\dot{\phi}}^2 \ll V$,
  \item[(iii)] $H \ll \Gamma$,
  \item[(iv)] $|\dot{\rho_{\rm r}}| \ll 4H\rho_{\rm r},\,\Gamma\dot{\phi}^2$,
  \item[(v)]  $|\ddot{\phi}| \ll (3H + \Gamma)\dot{\phi},\,V'(\phi)$.
\eed
The first two conditions are required in order to have inflation, the
third is the criterion for warm inflation as opposed to standard
inflation, without the fourth requirement it would be unreasonable to
assume that $\Gamma$ is constant, and the final criterion is the {\it
  slow rolling} condition for the inflaton dynamics. If these
conditions are satisfied, Eqs. (\ref{eq:eqn}) and
(\ref{eq:reqn}) reduce to
\bea
  \Gamma\dot{\phi} +V' \simeq 0 
    &\Longleftrightarrow&
  \dot{\phi} \simeq - \frac{V'}{\Gamma}
             \simeq - \frac{m^2 \phi}{\Gamma}, \\ 
  4H\rho_{\rm r} \simeq \Gamma\dot{\phi}^2
    &\Longleftrightarrow&
  \rho_{\rm r} \simeq \frac{\Gamma\dot{\phi}^2}{4H}
           \simeq \frac{V'^2}{4H\Gamma} 
           \simeq \sqrt{\frac38} \frac{m^3 M_{G} \phi}{\Gamma}, \\
    &\Longleftrightarrow&
  T_{\rm r} = \lmk \frac{30 \rho_{\rm r}}{g_{\ast} \pi^2} \rmk^{\frac14}
        \simeq \lmk \frac{675}{4 g_{\ast}^2 \pi^4} \rmk^{\frac18}   
               \lmk \frac{m^3 M_{G} \phi}{\Gamma} \rmk^{\frac14},
\eea
where $g_{\ast}$ is the number of the relativistic degrees of freedom.

Conditions (i) and (ii) imply the dominance of the vacuum energy. The
first is satisfied for
\beq
  \phi \gg \phi_{\rm end} \equiv \sqrt{\frac{3}{2}} \frac{m}{\Gamma} M_{G},
\eeq
and the second is then obeyed for all values of $\phi$ provided that
\beq 
m \ll \Gamma \, .
\eeq
Thus, warm inflation ends at $\phi = \phi_{\rm end}$. At that time,
the temperature of radiation becomes
\beq
  T_{\rm end} \equiv
  T_{\rm r}(\phi = \phi_{\rm end}) \sim m \sqrt{\frac{M_{G}}{\Gamma}}.
\eeq
The condition (iii) implies that the dominant friction term in the
inflaton equation of motion is given by the coupling to other
particles rather than by the Hubble expansion and is satisfied for
\beq
  \phi \ll \sqrt{6}\,\frac{\Gamma}{m} M_{G}.
\eeq
The condition (iv) implies the constancy of the energy density of
radiation and is valid as long as $\phi > \phi_{\rm end}$. The last
requirement is the so-called slow-roll condition and is also satisfied
if $m \ll \Gamma$. Combining these results, we conclude that warm
inflation takes place while $\phi$ is in the range given by
\beq \label{range}
\phi_{\rm end} \lesssim \phi \lesssim (\Gamma/m) M_{G} 
\eeq
provided that $m \ll \Gamma$.

The number $N(\phi)$ of e-foldings of inflation between when the
inflaton field has the value $\phi$ in the range given by
Eq. (\ref{range}) and when $\phi = \phi_{\rm end}$ can easily be
estimated and yields the following relation between $\phi_{N}$ (the
initial value of $\phi$ which gives N e-foldings) and $N$:
\beq
  N = \int H dt \simeq \frac{\Gamma}{\sqrt{6}\,m M_{G}}
                        (\phi_{N}-\phi_{\rm end}),
\eeq
with the result $\phi_{N} \simeq \sqrt{6}\,N m M_{G} / \Gamma$. Taking
the scale of the cosmic microwave background (CMB) anisotropies
measured by the Cosmic Background Explorer (COBE) satellite to
correspond to $N = 60$, then
\beq
\phi_{\rm COBE} = \phi_{60}
\sim 150\,m M_{G} / \Gamma \sim 150\,\phi_{\rm end} \,.
\eeq

For $T_{\rm r} > H$, which corresponds to $\phi <
(M_{G}/m\Gamma)^{1/3} M_{G}$, thermal fluctuations dominate over
quantum fluctuations \cite{thermal}.  As shown in the last
reference of \cite{principle} (see also \cite{thermal2}) the
root mean square of fluctuations of the inflaton is given by
\beq
  \la (\delta \phi)^2 \ra \simeq \frac{1}{2\pi^{2}} 
                                    \sqrt{\Gamma H} T_{\rm r}.
\eeq
Based on these initial conditions for fluctuations generated during
inflation,, the final\footnote{Final means when the scale reenters the
Hubble radius at late times.} amplitude of the curvature perturbation
$\Phi_{A}$ (the relativistic gravitational potential in longitudinal
gauge - see \cite{MFB}) on a comoving scale whose physical wavelength
equals the Hubble radius during the period of warm inflation at
$\phi=\phi_{N}$ is given by \cite{flucts1,flucts2,thermal2}
\beq
  \Phi_{A} \sim f H \frac{\sqrt{\la (\delta \phi)^2 \ra}}{\dot{\phi}} 
           \sim 0.02 \lmk \frac{\Gamma^9 \phi_{N}^3}{M_{G}^{9} m^{3}} 
                    \rmk^{\frac18}, 
\eeq
where $f=3/5~(2/3)$ in the matter (radiation) domination (this result
was derived using the full general relativistic theory \cite{MFB} of
linear cosmological fluctuations in \cite{flucts2}). The COBE
normalization of CMB anisotropies requires $\Phi_{A} \simeq 3\times
10^{-5}$ at $N\simeq 60$ \cite{COBE}. This leads to the requirement
\beq
  \Phi_{A}(N=60) \sim 0.1 \lmk \frac{\Gamma}{M_{G}} 
                      \rmk^{\frac34}
                 \sim 10^{-5}.
\eeq
which yields $\Gamma \sim 10^{13}$ GeV.

In addition, the spectral index $n_{s}$ can be estimated to be
\cite{thermal2,flucts2}
\bea
  n_{s} - 1 &=& \frac{\dot{\phi}}{H} \frac{d}{d\phi}(\ln \Phi_{A})
            \sim \frac{3\sqrt{6}}{8} \frac{m M_{G}}{\Gamma \phi_{N}}, \\ 
            &\sim& 0.006 \qquad \qquad \qquad \qquad \qquad \qquad
              {\rm for} \quad \phi = \phi_{COBE} \sim 
                  150 \lmk \frac{m M_{G}}{\Gamma} \rmk.
\eea

Even after warm inflation ends, the friction term in the equation of
motion of the inflaton $\phi$ is still large so that the inflaton
continues to slow-roll instead of oscillating coherently. Hence, the
discussion of spontaneous baryogenesis in the reheating stage done in
Ref. \cite{CK} does not directly apply to the case of warm inflation.
After warm inflation ends, the dynamics of the inflaton is given by
\beq
 \left\{
 \begin{array}{l}
  \phi = \phi_{\rm end} \exp \lhk 
            - \frac{m^2}{\Gamma}(t - t_{\rm end})
                             \rhk, \\
  \dot\phi = - \sqrt{\frac32} \frac{m^3}{\Gamma^2} M_{G}\,\exp \lhk 
            - \frac{m^2}{\Gamma}(t - t_{\rm end})
                             \rhk, 
  \label{eq:dotphi}
 \end{array}\right.
\eeq
with $t_{\rm end} \simeq \Gamma/m^2$.

\section{Baryo/leptogenesis in warm inflation}

\label{sec:baryogenesis}

In this subsection, we show that spontaneous baryo/leptogenesis can
easily be realized in warm inflation. As is mentioned briefly in the
Introduction, we assume that the inflaton couples derivatively to the
$B-L$ current via an interaction Lagrangian
\beq
  \CL_{\rm eff} = \frac{1}{M}\,\del_{\mu}\phi\,J_{B-L}^{\mu},
  \label{eq:derivative}
\eeq
where $M$ is the cutoff scale. Assuming that $\phi$ is homogeneous,
the above coupling becomes
\beq
  \CL_{\rm eff} = \frac{\dot\phi}{M}\,n_{B-L}
                = \frac{\dot\phi}{M}\,(n_{b-l} - n_{\overline{b-l}}) 
                = \mu(t)\,n_{B-L},
\eeq
with $\mu(t)$ defined as
\beq
  \mu(t) \equiv \frac{\dot\phi}{M} .
\eeq

In contrast to the standard inflation models, radiation always exists
in close to thermal equilibrium. Then, if the time derivative $\dot\phi$ is
effectively nonzero, $\mu(t)$ becomes the effective time-dependent
chemical potential, which induces the $B-L$ asymmetry even in thermal
equilibrium. Such a thermal equilibrium baryo/leptogenesis scenario is
discussed in the context of quintessence \cite{LFZ,FNT,Yamaguchi}.
However, for this mechanism to work in the context of warm inflation,
the decoupling temperature of the $B-L$ violating operator would have
to be fine-tuned to be equal to the temperature of the radiation at
the end of warm inflation. Obviously, it needs to be lower or equal -
else there would be no thermal equilibrium for baryo- or leptogenesis
to occur towards the end of the period of inflation. But the
decoupling temperature cannot be lower either.
Since $\dot\phi$ decays exponentially after warm inflation ends, then
if the decoupling temperature were lower than $T_{\rm end}$, there
would be a time interval after the end of warm inflation during which
$B-L$ violating processes would be in thermal equilibrium but the
chemical potential for $B-L$ number would be effectively zero, and
during which therefore the $B-L$ number density would be driven to
zero.  Because of this fine-tuning, we do not consider this
possibility any further.

In this paper, we consider another possibility, namely one in which
the $B-L$ asymmetry is generated dynamically like in the Affleck-Dine
baryogenesis scenario. In this case, the upper bound on the decoupling
temperature for $B-L$ violating processes no longer is present.
Taking the coupling (\ref{eq:derivative}) into account, the equation
of motion of the inflaton is changed to
\beq
 \ddot{\phi} + (3H + \Gamma)\dot{\phi} 
   - \frac{1}{M} (\dot{n}_{B-L} + 3 H n_{B-L}) + V'(\phi) = 0. 
 \label{eq:eom2}
\eeq
When the $B-L$ current is not conserved, the divergence of the current
does not disappear and is replaced by a current violating operator.
We simply assume that the $B-L$ current is not conserved and such an
operator just gives rise to an additional decay channel for the
inflaton. In fact, if such a derivative coupling is, for example,
derived from the Yukawa coupling or something like that, the inflaton
interacts with the particles with the $B-L$ charges, which causes the
violation of the $B-L$ current. As stated before and shown in
\cite{BGR,YL,principle}, the dissipative term of the inflaton has a
complicated form and is not necessarily local in time.  However, by
using the adiabatic-Markovian approximation \cite{BGR,YL,principle},
the dissipative term in the equation of motion of the inflaton can be
approximated as a local term which is proportional to the time
derivative of the inflaton. The coefficient of proportionality
$\Gamma_{B-L}$ is still complicated and depends on the form of the
interactions, but, for simplicity, we assume it is a constant. It is
straightforward to extend the analysis to the case of a coefficient
which depends on the field value.  Thus, the equation of motion of the
inflaton is changed to\footnote{Strictly speaking, $\Gamma_{B-L}$
should be included in $\Gamma$. However, since we want to pay special
attention to the term related to the $B-L$ current, we keep it
separate.}
\beq
 \ddot{\phi} + (3H + \Gamma + \Gamma_{B-L})\dot{\phi} 
      + V'(\phi) = 0. 
 \label{eq:eom3}
\eeq
If $\Gamma_{B-L}$ is much smaller than $\Gamma$, the dynamics of the
inflaton is not changed much. Comparing the two Eqs.
(\ref{eq:eom2}) and (\ref{eq:eom3})\footnote{As for this comparison,
  a subtlety was raised in Ref. \cite{DF}. While Eq. (\ref{eq:eom2})
  is an operator equation, Eq. (\ref{eq:eom3}) is obtained after
  vacuum averaging.  The authors of \cite{DF} showed that the average
  value $\la \dot{n}_{B-L} \ra$ is complicated and not given by the
  above simple comparison when the inflaton oscillates coherently.
  However, in our case, the inflaton does not oscillate coherently but
  slowly rolls so that this subtlety does not matter. In particular,
  for the values of the parameters which we use, the classical
  approximation is justified, and there is no {\it energy problem} as
  discussed in the second reference of \cite{DF}.}, it follows that
the time evolution of the $B-L$ number density is given by
\beq
  \dot{n}_{B-L} + 3 H n_{B-L} = 
    - M\, \Gamma_{B-L} \dot{\phi}.
\eeq

As given in Eq. (\ref{eq:dotphi}), $\dot\phi$ decays exponentially
after warm inflation so that, except for the dilution due to the
adiabatic expansion of the universe, $n_{B-L}$ changes significantly
only during the short period $\Delta t \simeq \Gamma/m^2$ after
inflation.  Then, the ratio between the $B-L$ number density $n_{B-L}$
and the entropy density $s = \frac{2\pi^2}{45} g_{\ast s} T^3$ can be
roughly estimated to be
\bea
  \frac{n_{B-L}}{s} &\simeq& 
     M\,\Gamma_{B-L} \frac{m^3}{\Gamma^2} M_{G}
     \Delta t \left/ \frac{2\pi^2}{45} g_{\ast s} T_{\rm end}^3
              \right. \non \\
                    &\simeq&
     0.02\,\frac{M \Gamma_{B-L}}{m^2}
          \lmk \frac{\Gamma}{M_{G}} \rmk^{\frac12}.
\eea
Here we took $g_{\ast s} \sim 100$. If $T_{\rm end}$ is higher than
the temperature of the electroweak phase transition, a part of the
$B-L$ asymmetry at that time is converted into the baryon asymmetry
through the sphaleron processes \cite{sphaleron}. Then, the
baryon-to-entropy is given by
\bea
  \frac{n_B}{s} &\simeq&
                  \frac{8}{23} \frac{n_{B-L}}{s} \non \\
                &\simeq& 
     0.01\,\frac{M \Gamma_{B-L}}{m^2}
          \lmk \frac{\Gamma}{M_{G}} \rmk^{\frac12} \non \\
                &\simeq& 
     3 \times 10^{-5}\,\frac{M \Gamma_{B-L}}{m^2},
\eea
where we have used $\Gamma/M_G \sim 10^{-5}$. If we take $M
\Gamma_{B-L} / m^2 \sim 3 \times 10^{-6}$, then $n_B/s \sim 10^{-10}$.
It is thus quite easily possible to obtain the observed baryon to entropy
ratio.

Finally, we comment on the decoupling temperature of the $B-L$
violating interactions. For example, we consider the following
Lagrangian for lepton number violation:
\begin{equation}
\label{eq:dim5}
{\cal L}_{\not{L}} = \frac{2}{v}\,l\, l\, H\, H +\, {\rm H.c.},
\end{equation}
where $v$ is the scale characterizing the interaction which can be related
to the heavy Majorana mass for the right-handed neutrino (in the
context of the see-saw mechanism) in the following way:
\begin{equation}
m_\nu = \frac{4 \la H \ra^2}{v}\,.
\end{equation}
In the above, $l$ and $H$ represent the left handed lepton doublet 
and the Higgs doublet, respectively. Then, the lepton number violating 
rate of this interaction is~\cite{sarkar}
\begin{equation}
\label{eq:l_vio_rate}
\Gamma_{\not{L}} \sim 0.04 \frac{T^3}{v^2}.
\end{equation}
The decoupling temperature is now calculated as
\begin{eqnarray}
\label{eq:decT}
T_D &\sim&3 \times 
10^{11} {\rm GeV} \left(\frac{v}{10^{14} {\rm GeV}}\right)^2,\non\\
   &\sim & 2 \times 10^{14} {\rm GeV} 
                     \left(\frac{m_\nu}{0.05{\rm eV}}\right)^{-2},
\end{eqnarray}
where we set the number of effective degrees of freedom for
relativistic particles to be $100$. On the other hand, taking $m \ll
\Gamma, \Gamma/M_G \sim 10^{-5}$ into account, we can obtain an upper
bound on $T_{\rm end}$ of the form
\beq
  T_{\rm end} \sim m \sqrt{\frac{M_G}{\Gamma}} 
    \ll 3 \times 10^{-8} M_G \sim 7 \times 10^{10} \,{\rm GeV}.
\eeq
Thus, the decoupling temperature $T_{D}$ is much higher than $T_{\rm
  end}$.

\section{Discussion and Conclusions}

\label{sec:con}

In this paper, we have discussed spontaneous baryo/leptogenesis in
warm inflation. Though radiation always exists in warm inflation,
spontaneous baryogenesis in thermal equilibrium does not work in
general because for such a baryogenesis mechanism to be successful
would require a fine-tuning of the decoupling temperature of the $B-L$
violating interaction. Instead, we considered another possibility, in
which the $B-L$ asymmetry is generated dynamically. In the standard
inflation models, the inflaton oscillates coherently after inflation.
On the other hand, in warm inflation, the inflaton still slowly rolls
even after inflation ends. We have shown that spontaneous baryogenesis
can be implemented rather easily in this situation. We have shown that
during the short period just after warm inflation ends, a sufficient
$B-L$ asymmetry can be generated to explain the presently observed
baryon asymmetry.

For successful warm inflation, the inflaton must couple to a very
large number of particles in order to maintain a large decay width.
This may be viewed as a disadvantage, but it also renders it rather
likely that interaction terms such as those which we postulate exist.
Note that some models for warm inflation motivated by string theory
have been proposed.  All we assumed in this paper is the existence of
a derivative coupling of the inflaton to the $B-L$ current. It would
be of interest to explore whether in the string-motivated models for
warm inflation such a derivative coupling can be accommodated given
the couplings of $\phi$ to a large number of other fields.

\subsection*{ACKNOWLEDGMENTS}

At Brown, this work is supported in part by the United States
Department of Energy under Contract DE-FG0291ER40688, Task A. M.Y. is
partially supported by the Japanese Grant-in-Aid for Scientific
Research from the Ministry of Education, Culture, Sports, Science, and
Technology.


\begin{thebibliography}{9}

\bib{inflation}
For example, A.D. Linde, {\it Particle Physics and Inflationary
Cosmology}, (Harwood, Chur, Switzerland, 1990).

\bib{warm}
A. Berera, 
\PRL{75}{3218}{95}; 
A.~Berera,
Phys.\ Rev.\ D {\bf 54}, 2519 (1996)
[arXiv:hep-th/9601134];\\
A.~Berera,
Phys.\ Rev.\ D {\bf 55}, 3346 (1997)
[arXiv:hep-ph/9612239].

\bib{prewarm}
I.~G.~Moss,
Phys.\ Lett. {\bf 154B}, 120 (1985);\\
J. Yokoyama and K. Maeda,
\PLB{207}{31}{88}; \\
R. Brout and P. Spindel,
\NPB{348}{405}{91}.

\bib{thermal}
A.~Berera and L.~Z.~Fang,
Phys.\ Rev.\ Lett.\  {\bf 74}, 1912 (1995)
[arXiv:astro-ph/9501024];\\
W.~L.~Lee and L.~Z.~Fang,
Int.\ J.\ Mod.\ Phys.\ D {\bf 6}, 305 (1997)
[arXiv:astro-ph/9706101].

\bib{BGR}
A.~Berera, M.~Gleiser, and R.~O.~Ramos,
Phys.\ Rev.\ D {\bf 58}, 123508 (1998)
[arXiv:hep-ph/9803394];\\
A.~Berera and R.~O.~Ramos,
Phys.\ Rev.\ D {\bf 63}, 103509 (2001)
[arXiv:hep-ph/0101049];\\
A.~Berera and R.~O.~Ramos,
arXiv:hep-ph/0210301.

\bib{YL}
J.~Yokoyama and A.~D.~Linde,
Phys.\ Rev.\ D {\bf 60}, 083509 (1999)
[arXiv:hep-ph/9809409].


\bib{principle}
A.~Berera, M.~Gleiser, and R.~O.~Ramos,
Phys.\ Rev.\ Lett.\  {\bf 83}, 264 (1999)
[arXiv:hep-ph/9809583];\\
A.~Berera and T.~W.~Kephart,
Phys.\ Rev.\ Lett.\  {\bf 83}, 1084 (1999)
[arXiv:hep-ph/9904410];\\
A.~Berera and T.~W.~Kephart,
Phys.\ Lett.\ B {\bf 456}, 135 (1999)
[arXiv:hep-ph/9811295];\\
A.~Berera,
Nucl.\ Phys. {\bf B585}, 666 (2000)
[arXiv:hep-ph/9904409].

\bib{DF}
A.~Dolgov and K.~Freese,
Phys.\ Rev.\ D {\bf 51}, 2693 (1995)
[arXiv:hep-ph/9410346];\\
A.~Dolgov, K.~Freese, R.~Rangarajan, and M.~Srednicki,
Phys.\ Rev.\ D {\bf 56}, 6155 (1997)
[arXiv:hep-ph/9610405].

\bib{CK}
A.~G.~Cohen and D.~B.~Kaplan,
Phys.\ Lett.\ B {\bf 199}, 251 (1987).

\bib{AD}
I.~Affleck and M.~Dine,
Nucl.\ Phys. {\bf B249}, 361 (1985).

\bib{thermal2}
A.~N.~Taylor and A.~Berera,
Phys.\ Rev.\ D {\bf 62}, 083517 (2000)
[arXiv:astro-ph/0006077].

\bib{MFB}
V.~F.~Mukhanov, H.~A.~Feldman, and R.~H.~Brandenberger,
Phys.\ Rep.\  {\bf 215}, 203 (1992).

\bib{flucts1}
W.~Lee and L.~Z.~Fang,
Phys.\ Rev.\ D {\bf 59}, 083503 (1999)
[arXiv:astro-ph/9901195].

\bib{flucts2}
H.~P.~De Oliveira and S.~E.~Joras,
Phys.\ Rev.\ D {\bf 64}, 063513 (2001)
[arXiv:gr-qc/0103089].

\bib{COBE}
C.~L.~Bennett {\it et al.},
Astrophys.\ J.\ Lett. {\bf 464}, L1 (1996)
[arXiv:astro-ph/9601067].

\bib{LFZ}
M.~Z.~Li, X.~L.~Wang, B.~Feng and X.~M.~Zhang,
Phys.\ Rev.\ D {\bf 65}, 103511 (2002)
[arXiv:hep-ph/0112069];\\
M.~Li and X.~Zhang,
arXiv:hep-ph/0209093.

\bib{FNT}
A.~De Felice, S.~Nasri, and M.~Trodden,
\PRDD{67}{043509}{03}
[arXiv:hep-ph/0207211].

\bib{Yamaguchi}
M.~Yamaguchi,
arXiv:hep-ph/0211163.

\bib{sphaleron}
V.~A.~Kuzmin, V.~A.~Rubakov and M.~E.~Shaposhnikov,
Phys.\ Lett. {\bf 155B}, 36 (1985);\\
S.~Y.~Khlebnikov and M.~E.~Shaposhnikov,
Nucl.\ Phys. {\bf B308}, 885 (1988).

\bibitem{sarkar}
U. Sarkar, 
arXiv:hep-ph/9809209;\\
W. Buchmuller, 
arXiv:hep-ph/0101102.

\end{thebibliography}
\end{document}